\documentclass[twocolumn,showpacs,preprintnumbers,prl,aps,amssymb,superscriptaddress]{revtex4}
\usepackage{graphicx}
\usepackage{amsmath}
\usepackage{dcolumn}
\usepackage{bm}
\usepackage{color}
\usepackage{verbatim}

\begin{document}

\newcommand{\Co}{CeCoIn$_5$}
\newcommand{\Rh}{CeRhIn$_5$}
\newcommand{\ie}{{\it i.e.}}
\newcommand{\eg}{{\it e.g.}}
\newcommand{\etal}{{\it et al.}}
\newcommand{\tc}{{$T_{\rm c}$}}
\newcommand{\TFL}{{$T_{\rm FL}$}}
\newcommand{\hc}{{$H_{\rm c2}$}}
\newcommand{\hstar}{{$H^{\star}$}}

\title{Quantum critical quasiparticle scattering within the superconducting state of CeCoIn$_5$}


\author{Johnpierre~Paglione}
\email{paglione@umd.edu}
\address{Center for Nanophysics and Advanced Materials, Department of Physics, University of Maryland, College Park, MD 20742}
\affiliation{Canadian Institute for Advanced Research, Toronto, Canada M5G 1Z8}

\author{M.~A.~Tanatar}
\affiliation{D\'epartement de physique \& RQMP, Universit\'e de Sherbrooke, Sherbrooke, Canada J1K 2R1} 
\affiliation{Ames Laboratory USDOE and Department of Physics and Astronomy, Iowa State University, Ames, Iowa 50011, USA}

\author{J.-Ph. Reid}
\affiliation{D\'epartement de physique \& RQMP, Universit\'e de Sherbrooke, Sherbrooke, Canada J1K 2R1} 

\author{H.~Shakeripour}
\affiliation{Department of Physics, Isfahan University of Technology, Isfahan 84156-83111, Iran} 

\author{C.~Petrovic}
\affiliation{Canadian Institute for Advanced Research, Toronto, Canada M5G 1Z8}
\affiliation{Department of Physics, Brookhaven National Laboratory, Upton, New York 11973}

\author{Louis~Taillefer}
\email{Louis.Taillefer@USherbrooke.ca}
\affiliation{Canadian Institute for Advanced Research, Toronto, Canada M5G 1Z8}
\affiliation{D\'epartement de physique \& RQMP, Universit\'e de Sherbrooke, Sherbrooke, Canada}

\date{\today}


\begin{abstract}

The thermal conductivity $\kappa$ of the heavy-fermion metal \Co~was measured in the normal and superconducting states as a function of temperature $T$ and magnetic field $H$,  for a current and field parallel to the $[100]$ direction.  Inside the superconducting state, when the field is lower than the upper critical field \hc,  $\kappa/T$ is found to increase as $T \to 0$, just as in a metal and in contrast to the behavior of all known superconductors. This is due to unpaired electrons on part of the Fermi surface, which dominate the transport above a certain field. The evolution of $\kappa/T$ with field reveals that the electron-electron scattering (or transport mass $m^\star$) of those unpaired electrons diverges as $H \to$~\hc~from below, in the same way that it does in the normal state as $H \to$~\hc~from above. This shows that the unpaired electrons sense the proximity of the field-tuned quantum critical point of \Co~at \hstar~=~\hc~even from inside the superconducting state. The fact that the quantum critical scattering of the unpaired electrons is much weaker than the average scattering of all electrons in the normal state reveals a $k$-space correlation between the strength of pairing and the strength of scattering, pointing to a common mechanism, presumably antiferromagnetic fluctuations.

\end{abstract}

\maketitle


With the discovery of iron-based superconductors \cite{FeAs}, the interplay of magnetism and superconductivity has become an increasingly important topic of condensed matter physics. The archetypal evidence of magnetically-mediated superconductivity in the heavy-fermion metal CeIn$_3$ \cite{Mathur} linked unconventional Cooper pairing with magnetic fluctuations emanating from a quantum critical point (QCP), a scenario widely believed to explain the common appearance of superconductivity in the vicinity of antiferromagnetic order in heavy fermion, organic, pnictide and cuprate families of superconductors \cite{scatterpair}. 

The heavy-fermion superconductor \Co\ \cite{Petrovic_Co} continues to receive considerable attention \cite{Allan,Zhou,Tokiwa,Howald}.  Low-temperature transport \cite{Paglione_QCP,Paglione_WFa} and specific heat \cite{Bianchi_QCP} studies have revealed a magnetic field-tuned QCP, with a critical field \hstar\ that anomalously coincides with \hc, the upper critical field for superconductivity. The pinning of \hstar\ to \hc\ was subsequently shown to hold regardless of field orientation \cite{Ronning} or suppression of the superconducting state by impurities \cite{Bauer}, suggesting a novel form of quantum criticality closely linked with the superconducting state.  Recent work has revealed other examples of systems that appear to have a field-tuned QCP pinned to \hc,  including cuprates \cite{Shibauchi_PNAS,Butch} and iron pnictides \cite{Dong}.

Together with large angle scattering evidence \cite{Paglione_WFa}, the presence of similar critical behavior in the ordered antiferromagnet \Rh\ under pressure \cite{Park_pss} strongly suggests that the QCP for  $H \parallel c$ configuration in \Co\ is also magnetic in nature, although magnetic order was not observed in muon spin rotation \cite{musR} or neutron scattering measurements \cite{neutrons}. However, for $H \parallel ab$ neutron scattering \cite{Kenzelmann,Das} and nuclear magnetic resonance \cite{Curro} measurements have found field-induced antiferromagnetism in the vicinity of \hc\, suggesting magnetism grows gradually with increasing field \cite{Morten,Lapertot}.  However, the relation between quantum criticality and superconductivity in \Co~remains elusive, in particular due to the strong first-order character of \hc\ below $T \simeq 1$~K \cite{Bianchi_first},  making a connection between \hc\ and \hstar\ unlikely. This raises the fundamental question of whether the fluctuations associated with the field-tuned QCP in \Co\ are in any way present in the superconducting state and involved in the pairing.


In this Letter, we show that quasiparticle heat transport in the superconducting state of CeCoIn$_5$ reflects the same quantum critical behavior that characterizes transport in the normal state.  This observation provides us with an opportunity to study the field-tuned QCP from both {\it below and above} \hc. We find a similarly rapid increase of the quasiparticle mass on tuning to \hc\ from either side, consistent with the existence of a singular and continuous critical point, despite the first-order transition.  We also find a ten-fold decrease in the inelastic scattering strength upon crossing \hstar\ into the superconducting state, proving a direct link between scattering and pairing, as the Fermi-surface regions of strongest scattering are also those that are most strongly gapped. We therefore infer that the antiferromagnetic fluctuations associated with the QCP in \Co\ are also involved in the pairing.


High-quality single crystals of \Co\ and CeIrIn$_5$ were grown by the self-flux method \cite{Petrovic_Co}, with superconducting transition temperatures \tc~=~2.3~K and \tc~=~0.4~K, respectively. Platelet-shaped samples with typical dimensions $\sim 2 \times 0.2 \times 0.05$~mm$^3$ were prepared for transport measurements along the [100] direction, using the same four-wire contacts for both electrical and thermal conductivity.  Thermal conductivity was measured with a one-heater, two-thermometer steady-state technique and in-situ thermometer calibration in high fields, using low-resistance indium solder contacts to avoid electron-phonon decoupling effects at low temperatures \cite{Smith,Tanatar_WF}, and heat currents applied along the [100] crystallographic direction and magnetic field along either [001] or [100], to within 1$^{\circ}$ alignment.


\begin{figure}[!]
   \resizebox{8cm}{!}{
 \includegraphics[width=8cm]{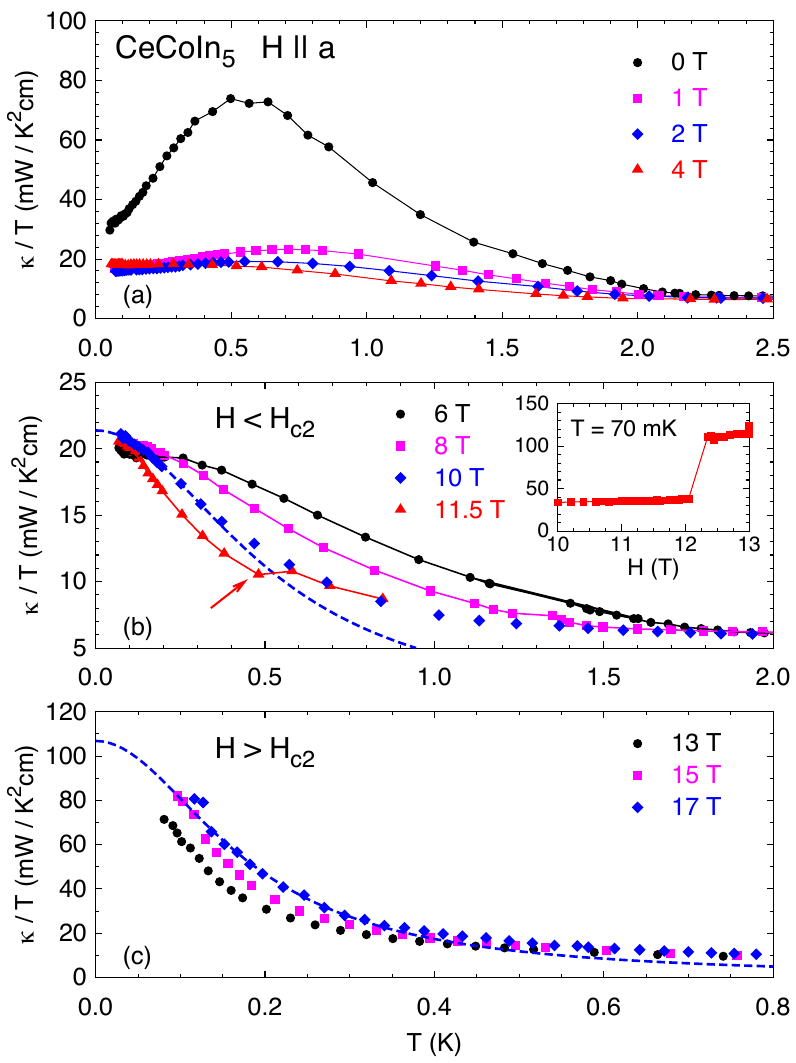}}
 \caption{\label{fig2} 
 In-plane thermal conductivity of \Co\ for $H\parallel~a$, plotted as $\kappa/T$ vs $T$. 
 (a) For $H < 5$~T, as indicated.
 (b)~For  5~T $< H <$~\hc, as indicated.
 {\it Inset}: 
 field dependence of $\kappa/T$ at $T = 70$~mK, showing the sharp first-order transition at \hc~=~12~T. 
The transition is also detected as a function of temperature, in the data at $H = 11.5$~T (red arrow, main panel).
 (c) For $H >$~\hc, in the normal state.
 The blue dashed lines in panels b) and c) are a fit of the data at $H = 10$~T  and $H = 17$~T, respectively, 
 to the Fermi-liquid expression, $\kappa/T = L_0 / (w_0 + BT^2)$, where $L_0 \equiv (\pi^2/3) (k_{\rm B}/e)^2$. 
  }
\end{figure}


In Fig.~\ref{fig2}, the electronic thermal conductivity $\kappa/T$ of \Co\ is presented for magnetic fields up to 17~T applied along the heat current ($H \parallel a$), covering the superconducting state below \hc~=~12~T and the normal state above \hc. 

There are two unusual features of \Co\ that must be born in mind. First, \Co\ is an extreme multi-band superconductor \cite{uncondensed}, in the sense that a tiny magnetic field (of order 10~mT~\cite{Seyfarth}) kills superconductivity on part of the Fermi surface, so that some of the carriers behave like normal-state quasiparticles even deep inside the superconducting state. These unpaired (uncondensed, ungapped) electrons dominate the thermal conductivity in the $T=0$ limit, and 90\% of the residual linear term $\kappa/T$ at $T \to 0$ is due to them, with only some 10\% coming from nodal quasiparticles \cite{uncondensed}. At intermediate temperatures, nodal quasiparticles become thermally excited and cause a peak in $\kappa/T$ vs $T$ (Fig.~\ref{fig2}a). However, applying a magnetic field introduces vortices that scatter these nodal quasiparticles and suppresses their contribution to $\kappa$ at all temperatures. As a result, for $H > 4$~T, $\kappa(T)$ is purely metallic-like, completely dominated by the unpaired electrons.  Indeed, as seen in Fig.~\ref{fig2}b, all curves with 4~T $< H <$~\hc~show Fermi-liquid behavior at low temperature.

Second, the transition out of the vortex state, from $H <$~\hc~to $H >$~\hc, is a pronounced first-order transition \cite{Bianchi_first}. This is readily seen in a field sweep at low temperature, as shown in the inset of Fig.~\ref{fig2}b, where $\kappa(H)$ undergoes a sudden jump at \hc~=~12~T.

\begin{figure}[!]
   \resizebox{8cm}{!}{
 \includegraphics[width=8cm]{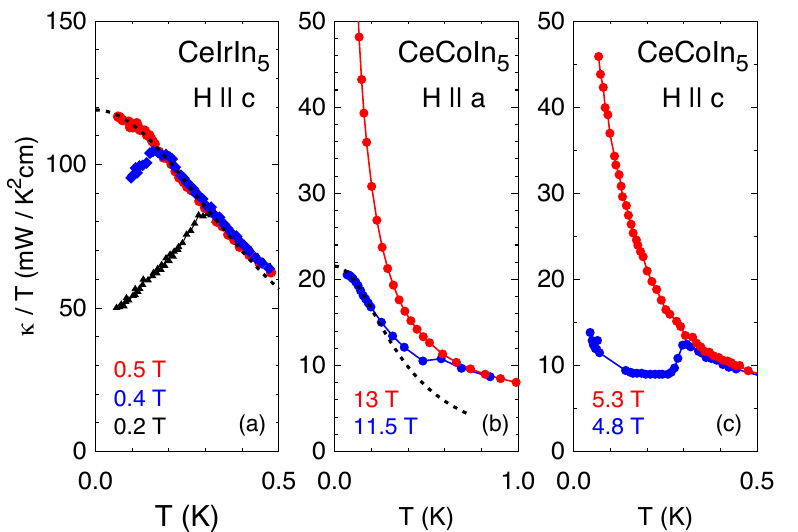}}
 \caption{\label{fig1} 
 (a) Thermal conductivity of CeIrIn$_5$, plotted as $\kappa/T$ vs $T$, for different values of the magnetic field as indicated, for $H \parallel c$.
For $H=H_{\rm c2}$ ($0.5$~T), $\kappa/T = L_0 / (w_0 + BT^2)$, as expected of a Fermi liquid (dotted line). 
For $H<H_{\rm c2}$ ($0.2$ and $0.4$~T), $\kappa/T$ decreases monotonically as $T \to 0$,
as found in most superconductors.  
 (b) Thermal conductivity of \Co\ in the superconducting (blue, $H =11.5$~T) and normal (red, $H=13$~T) states,
  for $H \parallel a$. 
In the normal state, $\kappa/T$ rises rapidly as $T \to 0$, a signature of the QCP at $H^\star = 12$~T,
with a Fermi-liquid regime observed only below 0.2~K.
 In the superconducting state, $\kappa/T$ rises as $T \to 0$, in stark contrast to the conventional behavior of CeIrIn$_5$.
 The conductivity mimics the behaviour of the normal state, showing that quantum criticality persists below \hc.
 The dotted line shows a fit to the Fermi-liquid function $L_0 / (w_0 + BT^2)$. 
 (c) Same as panel (b), but for $H \parallel c$, where $H^\star = 5$~T.
 }
 \end{figure}

We compare this to more conventional behavior observed in the closely-related superconductor CeIrIn$_5$, which unlike \Co\ has no field-tuned QCP \cite{Hamideh2016}, no small gap on part of its Fermi surface (hence no unpaired electrons at low field), and no first-order transition.
In Fig.~\ref{fig1}a, we show the thermal conductivity of CeIrIn$_5$, 
plotted as $\kappa/T$ vs $T$ \cite{hybrid,universalCeIrIn5}.
As in CeCoIn$_5$, the thermal conductivity of CeIrIn$_5$ is purely electronic, with negligible phonon contribution \cite{Paglione_WFa,Tanatar_WF,kasahara,hybrid,universalCeIrIn5,Hamideh2016}.
In the normal state, when $H =$~\hc~=~0.5~T or greater, $\kappa/T$ has the standard dependence of a Fermi liquid,
namely a thermal resistivity $w \equiv L_0 T / \kappa = w_0 + B T^2$, where $L_0 \equiv (\pi^2/3) (k_{\rm B}/e)^2$.
At $H=0$, $\kappa/T$ drops below $T_c$, and decreases monotonically to reach a non-zero residual value at $T=0$ \cite{hybrid,universalCeIrIn5},
the signature of nodes in the superconducting gap \cite{Shakeripour-NJP}.
The drop is simply due to a loss of thermally excited quasiparticles \cite{Graf}. 
It is in part compensated by a concomitant loss 
of electron-electron inelastic scattering, but in CeIrIn$_5$, this compensating effect is small,
since the strength of inelastic scattering at \tc~is only of order the elastic scattering, {\it i.e.} $B$\tc$^2 \simeq w_0$  \cite{hybrid,universalCeIrIn5}.
At intermediate fields ($0 < H <$~\hc), $\kappa/T$ continues to drop as $T \to 0$ (Fig.~\ref{fig1}a), again due to a loss of quasiparticle density.
The magnetic field also excites quasiparticles \cite{Vekhter}, in particular nodal quasiparticles at $T=0$, and hence increases $\kappa/T$ \cite{Shakeripour-NJP}.

As shown in Figs.~\ref{fig1}b and~\ref{fig1}c, the normal state inelastic scattering in \Co\ is completely different, and extremely strong, especially near the QCP in each field orientation (5~T for $H \parallel c$ and 12~T for $H \parallel a$).
For $H \parallel a$ (Fig.~\ref{fig1}b), $\kappa/T$ undergoes a ten-fold drop between $T=0$ and $T = 0.6$~K,
in the normal state at $H = 13$~T.
In the superconducting state at $H = 0$ (Fig.~\ref{fig2}a), 
$\kappa/T$ rises rapidly upon cooling below \tc~\cite{Movshovich,uncondensed}, because initially
the loss of inelastic scattering more than compensates for the loss of quasiparticles.
But eventually, at low temperature, $\kappa/T$ falls because of the decreasing quasiparticle density.

The resulting peak in $\kappa/T$ vs $T$ below \tc~is rapidly suppressed by a magnetic field (Fig.~\ref{fig2}a).
Above a certain field, namely when $H > 4$~T for $H \parallel a$, the fall at low temperature is no longer observed (Fig.~\ref{fig2}).
As seen in Fig.~\ref{fig1}b, at $H = 11.5$~T $<$~\hc, $\kappa/T$ shows no drop whatsoever as $T \to 0$, but rather exhibits the same $T$ dependence as the normal state, namely a Fermi-liquid behavior below 0.2~K, where $\kappa/T = L_0 / (w_0 + BT^2)$. 
This means that the heat carriers are not thermally excited, but simply unpaired (not gapped). 
Thanks to those unpaired electrons, the normal-state behavior of at least part of the Fermi surface can be studied {\it inside} the superconducting state, {\it below} the field-tuned QCP at $H^\star$.

\begin{figure}[!]
   \resizebox{8cm}{!}{
 \includegraphics[width=8cm]{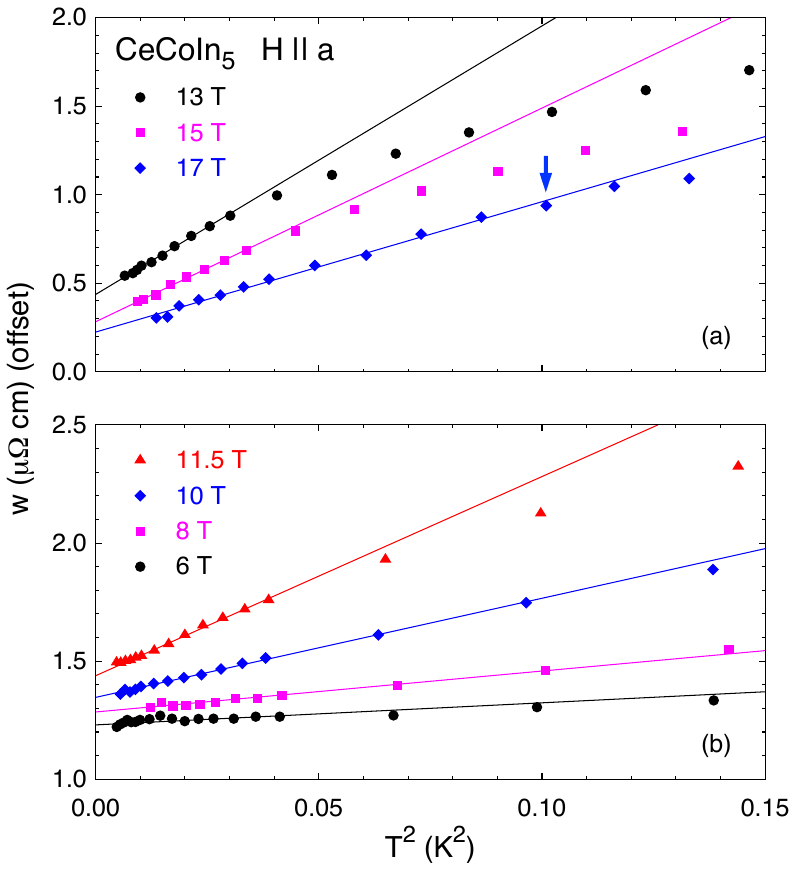}}
 \caption{\label{fig3} 
 In-plane thermal resistivity of \Co, defined as $w \equiv L_0T/\kappa$ and plotted vs $T^2$, for $H \parallel a$.
(a)
For $H >$~\hc, in the normal state.
(b)
For $H <$~\hc, in the superconducting state.
The solid lines are a fit of the data to the Fermi-liquid expression $w(T) = w_0 + BT^2$, where $w_0$ is the residual resistivity
due to elastic scattering and $B$ is the strength of the inelastic electron-electron scattering.
The fits are limited to an interval between $T=0$ and $T =$~\TFL~(arrow).
The fit parameters $w_0$,  $B$ and \TFL~are plotted in Fig.~\ref{fig4}.}
\end{figure}

\begin{figure}[!]
  \resizebox{8cm}{!}{
 \includegraphics[width=8cm]{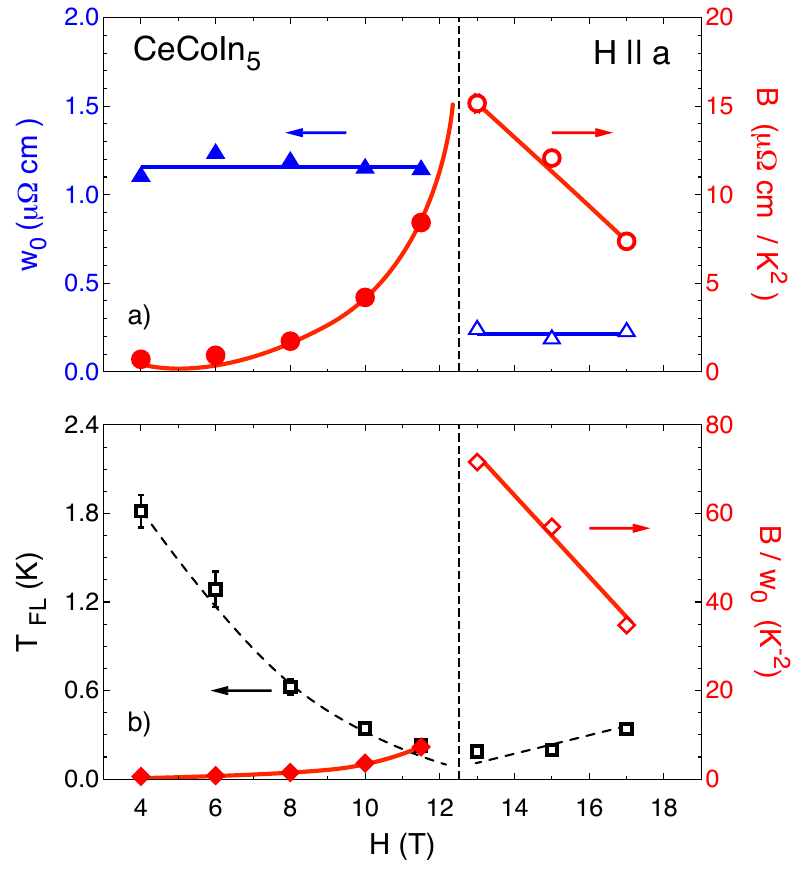}}
 \caption{\label{fig4} 
 Parameters obtained from a fit of the thermal resistivity $w(T)$ of \Co\ to the Fermi-liquid expression $w(T)=w_0 + BT^2$,
 as a function of magnetic field ($H \parallel a$).
The same fitting procedure, shown in Fig.~\ref{fig3}, is applied above \hc~=~12~T (vertical dashed line) and below \hc.
(a) Parameters $w_0$~(blue triangles, left axis), the residual thermal resistivity due to elastic scattering, and $B$ (red circles, right axis), due to inelastic electron-electron scattering, for both $H <$~\hc~(full symbols) and $H >$~\hc~(open symbols).
(b) Temperature \TFL~(black squares, left axis), the upper limit of the fitting interval (see Fig.~\ref{fig3}). Ratio $B / w_0$ (red diamonds, right axis), a measure of the strength of inelastic electron-electron scattering
relative to the strength of elastic scattering, for both $H <$~\hc~(full diamonds) and $H >$~\hc~(open diamonds). We see that $B / w_0$ undergoes a ten-fold drop upon entry into the superconducting state. All lines are a guide to the eye.}
\end{figure}

We demonstrate this by plotting the thermal resistivity $w(T)$ in Fig.~\ref{fig3} both above and below \hc, which is well described in the $T\to 0$ limit by the standard Fermi-liquid behavior,
$w(T) = w_0 + BT^2$,
with residual elastic scattering term $w_0$ and inelastic electron-electron scattering strength $B$.  Note that the extent of the $T^2$ regime, ending at $T =$~\TFL, changes as a function of field. In Fig.~\ref{fig4}, we plot $w_0$, $B$ and \TFL~vs $H$.

Let us first discuss the normal state, for $H >$~\hc.
In Fermi-liquid theory, the Wiedemann-Franz law requires that $w_0 = \rho_0$ (as observed~\cite{Paglione_WFa}), and the $T^2$ coefficient in $w(T)$ is proportional to the coefficient $A$ in the electrical resistivity $\rho(T) = \rho_0 + AT^2$, and typically $B~\simeq~2A$ \cite{Paglione_Rh}. 
Both $A$ and $B$ are related to the effective mass of the electrons, and $A \sim B \sim (m^*)^2$. 
The rapid rise of $B$ on approaching \hstar\ from above (Fig.~\ref{fig4})
is a signature of the field-tuned QCP, analogous to the 
rise of the $A$ coefficient of the resistivity for $H \parallel a$ \cite{Ronning}. (Note, that $A(H)$ dependence in our measurements matches very well with previous studies \cite{Ronning}.) In the local quantum criticality model, where fluctuations affect the entire Fermi surface \cite{QSi}, this rise is expected to follow $(H-H_{c})^{-1}$. For the spin-density wave scenario with only hot spot fluctuations, the field dependence becomes milder\cite{Hertz,Millis}.
The same parallel rise of $A$ and $B$ was previously reported for CeCoIn$_5$ in configuration $H \parallel c$~\cite{Paglione_WFa}. Nevertheless, the $A(H)$ (and similarly $B(H)$) dependence does not follow expectation of any theory, with specific heat, $\gamma (T) \equiv C/T$ revealing downward deviation from logarithmic divergence and simultaneous directional Wiedemann-Franz law violation for $H < $8~T \cite{Bianchi_QCP,Tanatar_WF}.
The fact that the residual resistivity $w_0$ is independent of $H$ (Fig.~\ref{fig4}) simply means that there is negligible magnetoresistance,
not surprisingly given the longitudinal configuration where current and field are parallel.

Let us now turn to the superconducting state, with $H <$~\hc.
The fact that $w_0$ is still nearly independent of $H$ (Fig.~\ref{fig4}a) is consistent with our interpretation that heat transport below \hc~is dominated by unpaired
electrons with metallic-like behavior, again with negligible magnetoresistance, at least for $H > 4$~T.
This is to be compared with the classic multi-band superconductor MgB$_2$, where a moderate in-plane field easily kills superconductivity on the small-gap quasi-3D $\pi$ Fermi surface, driving it into a gapless regime \cite{Gorkov}, but has little effect in exciting quasiparticles on the large-gap quasi-2D $\sigma$ Fermi surface \cite{Sologubenko}.
As a result, for $H \perp c$, $\kappa/T$ vs $H$ is nearly independent of $H$ above $\sim$\hc/10 and entirely due to the unpaired electrons on the $\pi$ surface for a wide range of fields.
In other words, just as in \Co, the unpaired electrons in MgB$_2$ completely dominate $\kappa$ inside the superconducting state
and allow one to probe the metallic state below \hc.

For \Co, this means we can directly study the inelastic scattering of unpaired electrons below \hc: the fact that $B$ rises rapidly upon approaching \hc\ from below (Fig.~\ref{fig4}a) provides direct evidence for the continuous nature 
of the field-tuned QCP in \Co, confirming that it survives the first-order superconducting transition. 
The unpaired electrons in the superconducting state clearly sense the presence of a QCP at $H=$~\hstar, with \hstar~$\simeq$~\hc, reminiscent of the mass divergence observed inside the superconducting state in BaFe$_2$(As$_{1-x}$P$_x$)$_2$ on both sides of the antiferromagnetic QCP \cite{P-Ba122}.

To compare the strength of inelastic scattering on either side of \hstar, we must first account for the large drop in carrier density as $H$ crosses below \hc.
A measure of this is provided by $w_0$, which is constant on either side of \hc,
but a factor of 6 larger below \hc~(Fig.~\ref{fig4}a). 
We infer that the carrier density (or spectral weight) of the unpaired electrons 
below \hc~is 6 times lower than that of the full Fermi surface above \hc.
To provide a meaningful measure of the strength of inelastic scattering, we therefore plot the ratio $B / w_0$ in Fig.~\ref{fig4}b, which drops abruptly by a factor 10 upon crossing below \hc. In other words, the unpaired electrons that prevail in the superconducting state experience a scattering that is {\it ten times weaker} than the average electron in the normal state just above \hc.
This reveals a powerful correlation between scattering and pairing:
those regions of the Fermi surface that experience a dramatically weaker inelastic 
scattering are the same that end up having the smallest gap, suggesting that heaviest carriers belonging to $\alpha$ and $\beta$ sheets of the Fermi surface \cite{Onuki,Julian} are most important for superconductivity, in accordance with the conclusion from STS measurements \cite{Allan}.

In summary, a continuous divergence exists in the electron-electron scattering of unpaired quasi-particles in CeCoIn$_5$ upon approach to the field-tuned QCP from both above and below the critical field, and that the amplitude of critical scattering is strongly suppressed in the superconducting state.
We conclude that the fluctuations associated with the QCP are responsible not only for scattering the electrons above and below \hc, but also for pairing these electrons, in what must be a strongly $k$-dependent fashion. 
This is reminiscent of the correlation between quantum critical scattering and pairing reported in organic \cite{Doiron_Tlinear},
pnictide \cite{Doiron_Tlinear} and cuprate superconductors \cite{scatterpair,Jin_LCCO}, whereby the strength of the linear-$T$ resistivity scales with \tc.
Moreover, in the single-band overdoped cuprate Tl-2201, the inelastic scattering was shown to be strongest in the same $k$-space regions
where the $d$-wave gap is maximal \cite{Hussey}. Similar ideas are discussed recently in relation to all unconventional superconductors \cite{Seamus_PNAS,JPHu}.


This work was supported by the Canadian Institute for Advanced Research and a Canada Research Chair (L.T.), and funded by NSERC, FRQNT and CFI. 
Work at the University of Maryland was supported by NSF-CAREER Grant No. DMR-0952716.
Part of the work was carried out at the Brookhaven National Laboratory, which is operated for the US Department of Energy by Brookhaven Science Associates (DE-Ac02-98CH10886) and in the Ames Laboratory, supported by the US Department of Energy, Office of Basic Energy Sciences, Division ofMaterials Sciences and Engineering, under Contract No. DE-AC02-07CH11358.


\end{document}